# Enhancing interfacial thermal conductance in Si/Diamond heterostructures by phonon bridge

Ershuai Yin, Qiang Li[*], Wenzhu Luo, Lei Wang

[a] MIIT Key Laboratory of Thermal Control of Electronic Equipment, School of Energy and Power Engineering, Nanjing University of Science & Technology, Nanjing, Jiangsu 210094, China

**Abstract:** This study investigates the mechanism of enhancing interfacial thermal transport performance in Silicon/Diamond (Si/Diamond) heterostructures using the phonon bridge. A heat transfer model for three-layer heterostructures is developed by combining First-principles calculations with the Monte Carlo method. The temperature distribution, spectral heat conductance, and interfacial thermal conductance are compared for Si/Diamond heterostructures with and without a silicon carbide (SiC) interlayer. The results show that the SiC interlayer effectively bridges low-frequency phonons in Si with mid-to-high-frequency phonons in Diamond, which forms a specific phonon bridge, significantly improving interfacial phonon transport. The influence of SiC interlayer thickness is further studied, revealing a size-dependent phonon bridge enhancement. For thin interlayers, intensified phonon boundary scattering weakens the bridging effect. Conversely, excessively thick interlayers increase the bulk thermal resistance, reducing overall interfacial thermal conductance. Thus, an optimal interlayer thickness exists, identified as 40 nm for SiC. Thirteen candidate interlayer materials, including SiC, AlN, α-$Si_3N_4$, β-$Si_3N_4$, and $Al_xGa_{1-x}N$ (x ranges from 0.1 to 0.9), are compared at various thicknesses. SiC emerges as the most effective interlayer material, increasing interfacial thermal conductance by 46.6% compared to the bilayer heterostructure. AlN ranks second, improving thermal conductance by 21.9%. These findings provide essential insights into the phonon bridge mechanism at heterogeneous interface thermal transport and offer valuable theoretical guidance for designing heterostructures with enhanced thermal transport performance.

**Keywords:** Interfacial thermal conductance; phonon bridge; Si/Diamond interface; thermal management

---

[*] Corresponding author. liqiang@njust.edu.cn (Q. Li)

# 1 Introduction

The interfacial thermal resistance (ITR) primarily arises from the mismatch in phonon properties between two dissimilar materials, which leads to phonon scattering and reflection at the interface, thereby impeding efficient cross-interface heat transport [1]. In recent years, with the widespread adoption of high thermal conductivity materials and substrate thinning technologies in chip fabrication, ITR has emerged as one of the main limitations to internal heat conduction in chips [2]. For instance, in a AlGaN/GaN-on-Diamond transistor, the temperature rise induced by the ITR between the GaN layer and the diamond substrate accounts for more than 40% of the total temperature rise [3]. Therefore, reducing ITC and enhancing interfacial thermal transfer performance have become pressing challenges in the electronic device thermal management [4].

Currently, the main strategies for enhancing heterogeneous interface thermal transport fall into two categories: increasing the contact area and introducing interlayers [5]. The contact area enhancement approach typically involves the construction of interfacial nanostructures to enlarge the actual contact region, thereby improving the interfacial thermal conductance (ITC). This method has been experimentally validated [6]. For example, Cheng et al. [7] fabricated ~100 nm trapezoidal structures on the Si surface and subsequently grew diamond films on top. Their measured results showed that the ITC of the nanostructured Si/Diamond interface reached 105 $Wm^{-2}K^{-1}$, representing an approximate 65% increase compared to the planar interface value of 63.6 $Wm^{-2}K^{-1}$. In addition, several theoretical studies have studied the enhancement mechanisms of interface nanostructures [8]. Our previous investigations [9][9,10] revealed that the interface nanostructures form enclosed cavities, in which phonons undergo multiple reflections and may transmit along the sidewalls, thus establishing an additional heat transfer pathway and significantly enhancing ITC. However, this approach suffers from limitations such as fabrication complexity, high cost, and a tendency to induce interfacial defects, all of which contribute to the poor controllability of the enhancement effect [11].

The introducing interlayer approach primarily involves introducing materials such as AlN or $Al_xGa_{1-x}N$ between the chip and the substrate to mitigate interfacial defects caused by lattice mismatch, including voids, cracks, and polycrystalline grain boundaries [12]. For example, Wang et al. [13] and Li et al. [14] independently investigated the effects of the AlN interlayer thickness on the interfacial thermal transport performance of GaN/SiC heterostructures. Their results indicated that the incorporation of an AlN interlayer significantly improved the interface crystalline quality, and that the ITC increased with increasing AlN thickness. The highest ITC observed experimentally was 372 $Wm^{-2}K^{-1}$, which remains considerably lower than the theoretically predicted value of the GaN/SiC interface (~505 $Wm^{-2}K^{-1}$). This discrepancy is mainly attributed to the fact that, although the interlayer improves interfacial quality, it simultaneously introduces an additional interfacial thermal resistance as well as an interlayer bulk thermal resistance. As a result, the enhancement in ITC is generally not achieved in most cases [5]. Similar findings have also been reported in other experimental studies [15].

To achieve the ITC that exceeds the theoretical limit, it is essential to address the mismatch in phonon frequency distribution between contacting materials, which prevents elastic phonons from transmitting across the interface. This requires the construction of new phonon transport pathways [16]. Some studies have demonstrated that designing interlayer to bridge the non-overlapping regions of the phonon spectrum between the two materials can enable a broader range of phonons to participate in interfacial heat transfer, which is known as phonon bridge [17]. For instance, Lee et al. [18] employed non-equilibrium molecular dynamics (NEMD) simulations using a unified Tersoff potential to investigate the effect of interlayer atomic mass on thermal transport across the GaN/SiC interface. Their results showed that varying the atomic mass of the interlayer effectively modulated the frequency distribution of optical phonons, thereby increasing the spectral overlap between GaN and SiC. Under the assumption of mass-only effects, the ITC was enhanced by approximately 27%. However, because the same Tersoff potential was used for all interlayer materials, the simulated behavior may deviate from real interfaces. Additionally, English et al. [19] and Polanco et al. [16] conducted NEMD simulations based on the Lennard-Jones potential to examine the influence of interlayer atomic mass on ITC. Their results revealed that, compared to structures without an interlayer, the ITC could be enhanced by up to 53%. This improvement was attributed to the interlayer's ability to effectively match the phonon spectrum of the adjoining materials.

At present, research on enhancing interfacial heat transfer through the construction of phonon bridges remains limited and largely relies on idealized assumptions, such as considering only the effect of interlayer atomic mass [5]. The main reason is the lack of effective methodologies for investigating how parameters such as interlayer material and thickness influence the ITC, making it difficult to accurately evaluate the enhancement effect. Existing studies on heterogeneous interfacial thermal transport are predominantly based on NEMD simulations, whose accuracy is highly dependent on the reliability of the employed potential function. However, the limited availability of potential function capable of accurately describing interactions across a wide range of materials severely constrains further research in this area [20]. Moreover, commonly used interlayer materials such as AlN and $Al_xGa_{1-x}N$ are, in most cases, not theoretically capable of enhancing interfacial heat transfer and thus may fail to realize the phonon bridge effect. Meanwhile, the influence of interlayer thickness on ITC may also be significant, yet studies remain scarce.

Therefore, this paper aims to explore the mechanism of enhancing heat transfer at heterogeneous interfaces by phonon bridge, using Si/Diamond heterogeneous interfaces as an example. A heat transfer model for three-layer heterostructures is developed by combining First-principles calculations with the Monte Carlo method. The temperature distribution, spectral heat conductance, and interfacial thermal conductance are compared for Si/Diamond heterostructures with and without a SiC interlayer. The influence of SiC interlayer thickness on the interfacial thermal conductance is further studied to explore the size dependence of the phonon bridge enhancement effect. Moreover, thirteen candidate interlayer materials, including SiC, AlN, α-$Si_3N_4$, β-$Si_3N_4$, and $Al_xGa_{1-x}N$ (x ranges from 0.1 to 0.9), are compared at various thicknesses.

## 2 Methodology

Figure 1(a) illustrates the schematic diagram of the Si/interlayer/Diamond three-layer heterostructures, while Figure 1(b) shows the Si/Diamond two-layer structure. The total width and height of the structure are $W$ and $H$, respectively, and the thickness of the interlayer is $h$. This study investigates the effects of various interlayer materials (e.g., SiC, AlN) and thicknesses on the Si/interlayer/diamond heterostructure ITC, in comparison with the ITC of the Si/Diamond structure without an interlayer. Thus, $h$ is the most critical parameter and varies from 1 nm to 250 nm. In the three-layer structure, the thicknesses of the Si and diamond layers are fixed at 100 nm each. As $h$ increases, the total height $H$ increases accordingly. The Si and diamond layers in the Si/Diamond two-layer structure are also set to 100 nm. The width $W$ and depth $D$ of the structure are both set to 100 nm. Periodic boundary conditions are applied in the $x$- and $z$-directions to eliminate the effect of size on results. A high temperature of 303 K is imposed at the top ($y = H$), while the bottom ($y = 0$) is maintained at 297 K, establishing a steady-state heat flux from top to bottom within the structure.

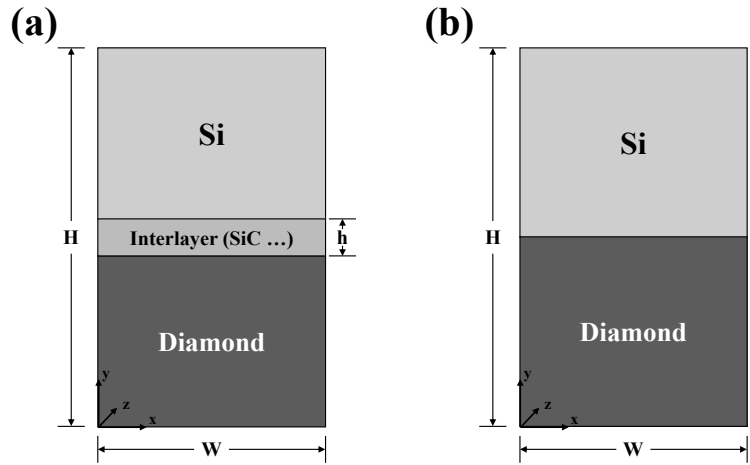

Fig. 1. (a) The schematic diagram of the Si/interlayer/diamond three-layer heterostructure (b) the Si/Diamond two-layer heterostructure.

Phonon transport within heterostructure is described by the deviational energy-based Boltzmann Transport Equation (BTE) under the relaxation time approximation (RTA) [21]:

$$\frac{\partial e^d}{\partial t} + \mathbf{v}_g(k,p) \cdot \nabla e^d = \frac{\left(e^{loc} - e^{eq}_{T_{eq}}\right) - e^d}{\tau(k,p,T)} \quad (1)$$

where $e^d = \hbar\omega(k,p)\left(f - f_{T_{eq}}^{eq}\right)$ is the deviational energy distribution, while $e^{loc}$ and $e_{T_{eq}}^{eq}$ represent the local and equilibrium energy distributions. $\omega(k,p), \mathbf{v}_g(k,p)$ and $\tau(k,p)$ correspond to the angular frequency, group velocity, and relaxation time. $k$ is the phonon wave vector, and $p$ denotes the phonon polarization. $f = \left(\exp(\frac{\hbar\omega(k,p)}{k_B T}) - 1\right)^{-1}$ is the Bose–Einstein distribution function at temperature $T$. $f_{T_{eq}}^{eq}$ is the Bose–Einstein distribution function at the equilibrium temperature $T_{eq}$.

The above deviational energy-based Boltzmann Transport Equation can be solved using the variance-reduced Monte Carlo method, the fundamental principles can be found in References [22,23]. In the previous work, we developed a numerical solution method for analyzing interfacial thermal transport across two-layer heterostructures containing nanostructures [9]. The main contribution of this work is the numerical solution method to more complex three-layer heterostructure. The challenge lies in accurately treating phonon interface interactions. When phonons propagate to an interface, they may either transmit through the interface or be reflected. In this work, for both internal interfaces in the three-layer heterostructure, the Diffuse Mismatch Model (DMM) is employed to evaluate the phonon transmission coefficient as a function of frequency [24]:

$$\tau_{1\rightarrow 2}(\omega') = \frac{\Delta V_2 \sum_{\mathbf{k},p} |\mathbf{v}_{g,2} \cdot \mathbf{n}| \delta(\omega'-\omega)}{\Delta V_1 \sum_{\mathbf{k},p} |\mathbf{v}_{g,1} \cdot \mathbf{n}| \delta(\omega'-\omega) + \Delta V_2 \sum_{\mathbf{k},p} |\mathbf{v}_{g,2} \cdot \mathbf{n}| \delta(\omega'-\omega)} \quad (2)$$

where $\Delta V$ represents the volumes of the discretized cells corresponding to the Brillouin zones, $\mathbf{v}_g$ is the phonon group velocity, and $\delta$ denotes the Dirac delta function.

Since phonon transport and interfacial transmission are highly dependent on the phonon properties, the accuracy of this method strongly relies on the input phonon parameters. Unlike previous studies that often adopted semi-empirical models [25,26] or gray-body approximations [8], this work obtains the full phonon dispersion and relaxation times directly from first-principles calculations, thereby avoiding errors arising from oversimplified phonon dispersion descriptions [27]. The computational procedure of phonon parameters is as follows: the second-order and third-order force constants for each material are first extracted using the VASP [28,29], PHONOPY [30], and thirdorder.py [31] packages. These force constants are then used in the almaBTE package [32] to compute the relevant phonon properties. The calculated phonon parameters include angular frequency, group velocity, relaxation time, and volumetric specific heat.

During the first-principles calculations, the local density approximation (LDA) functional is employed, with a plane-wave energy cutoff (ENCUT) set to 520 eV and the precision level set to High. The supercell sizes used for Si, diamond, SiC, AlN, GaN, α-Si$_3$N$_4$, and β-Si$_3$N$_4$ are 5×5×5, 6×6×6, 4×4×4, 5×5×5, 5×5×5, 2×2×2, and 2×2×4, respectively. During the third-order force constant calculations, the cutoff is set to the 5th or 6th nearest neighbors, depending on the crystal symmetry and atom number in the primitive cell, to ensure a balance between accuracy and computational efficiency. A 15×15×15 phonon **q**-point mesh is uniformly adopted for solving the BTE using the almaBTE package. For the $Al_xGa_{1-x}N$ alloy, a mixed crystal model is constructed via interpolation of the force constants between AlN and GaN, and its phonon properties at different compositions and temperatures are subsequently calculated using the almaBTE package. In the Monte Carlo simulation, the total number of phonons is set to $6×10^5$, and the equilibrium temperature is set at 300 K. The computational domain is discretized into a grid of 20×100×20 in the x, y, and z directions. For the Monte Carlo method, the grid is only used to accumulate phonons contributions to local temperature and heat flux bashe on the phonon trajectories. Therefore, an overly fine mesh is not necessary.

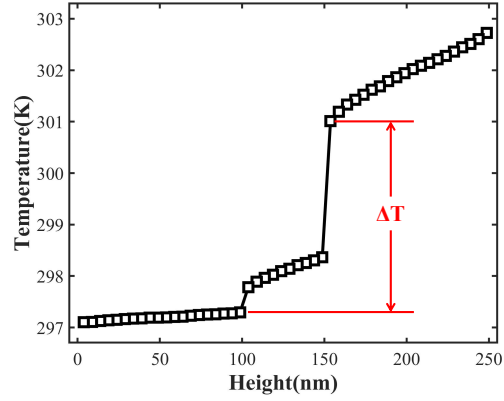

Fig. 2. Temperature distribution of the Si/SiC/diamond three-layer heterostructure with a 50 nm SiC interlayer.

After completing the Monte Carlo simulation, the temperature and heat flux distributions within the heterostructure can be obtained. The interfacial thermal conductance can then be calculated using the following equation [33,34]:

$$G = \frac{Q}{\Delta T} \qquad (3)$$

where $Q$ represents the steady-state heat flux across the interface, and $\Delta T$ denotes the temperature drop at the interface. For the three-layer heterostructure, a typical temperature distribution is shown in Figure 2. The average interfacial temperature drop is defined as the total temperature drop from material 1 to material 3, which includes the temperature drops caused by the two contact interfaces and the interlayer bulk thermal resistance.

# 3 Results and Discussion

## 3.1 Model verification

Figures 3(a), 3(b), and 3(c) demonstrate the thermal conductivities of Diamond, Si, SiC, AlN, $Al_xGa_{1-x}N$, $\alpha$-$Si_3N_4$, and $\beta$-$Si_3N_4$ calculated from first principles BTE method and compare them with available experimental or theoretical results reported in the literature. As shown in Figure 3(a), for materials such as diamond, Si, SiC, and AlN, the calculated thermal conductivities at various temperatures exhibit good agreement with experimental data, validating the high accuracy of the adopted computational approach. Figure 3(b) displays the thermal conductivities of $Al_xGa_{1-x}N$ alloys with different Al compositions, comparing the present results with both experimental data and those obtained from the Callaway model. The calculated values lie within the range of experimental results and show excellent consistency with the Callaway model. Figure 3(c) shows the calculated thermal conductivities of $\alpha$-$Si_3N_4$ and $\beta$-$Si_3N_4$ along the c-axis at 300 K, which also agree well with previous experimental and theoretical results, further demonstrating the reliability of the computational approach.

Figure 3(d) presents the computed ITC for the Si/Diamond, Si/SiC, and SiC/Diamond interfaces, along with comparisons to experimental values and results from other methods. The ITC obtained using the Monte Carlo method closely match those predicted by the DMM method, as both approaches incorporate full phonon dispersion and frequency-dependent transmission coefficients. The ITC for the Si/Diamond and Si/SiC interfaces obtained in this work are significantly lower than those predicted by empirical potential-based molecular dynamics simulations and are more consistent with experimental data. These results indicate that the computational methods developed in this work provides high accuracy and is suitable for investigations of interfacial thermal transport in heterostructures.

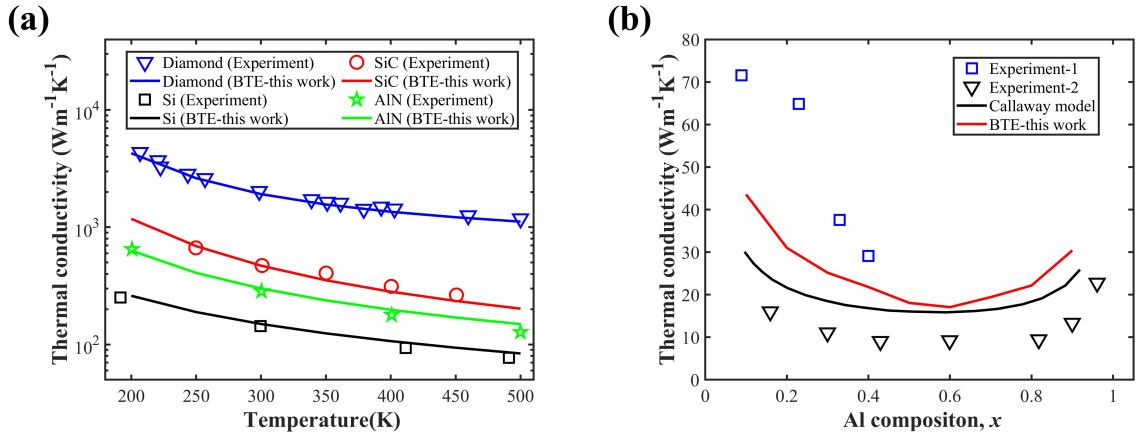

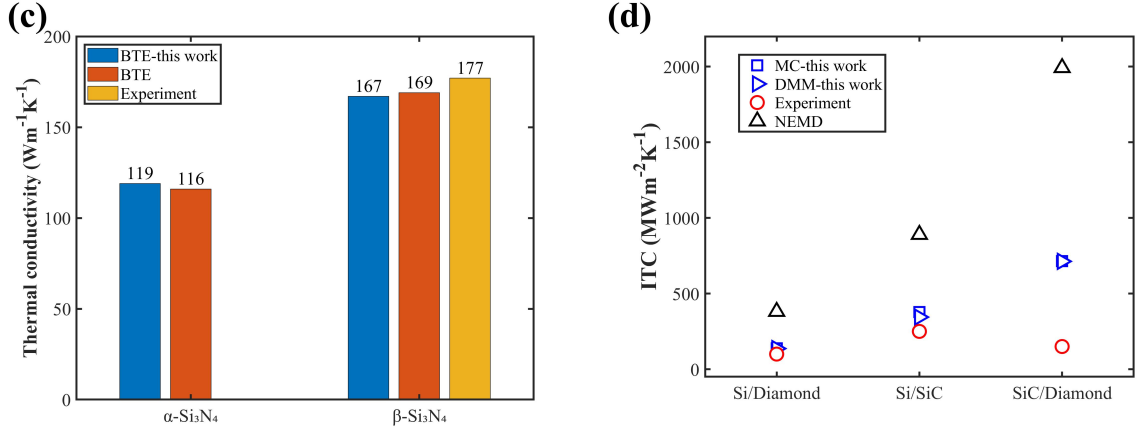

Fig.3. Model validation (a) Thermal conductivities of Diamond, Si, SiC, and AlN at different temperatures, where the experimental thermal conductivities of Diamond Si, SiC, and AlN are obtained from references [35], [36], [37], and [38]. (b) Thermal conductivity of $Al_xGa_{1-x}N$, where experimental and Callaway model results are from references [39,40]. (c) Thermal conductivities of α-$Si_3N_4$, and β-$Si_3N_4$ at 300 K along the c-axis, where the BTE and experimental results are from reference [41]. (d) Comparison of calculated and measured ITC at the Si/Diamond, Si/SiC, and SiC/Diamond interfaces, where the NEMD and experimental results are from references [5,42].

*3.2 Interfacial Thermal Transport in Si/Diamond Two-layer Heterostructures*

The interfacial thermal transport performance in Si/Diamond two-layer heterostructures is first studied, as ilustarted in Figure 4. Figure 4(a) illustrates the steady-state temperature distribution along the height direction. A distinct temperature discontinuity is observed at the interface, with a temperature drop of approximately 4.0 K, accounting for 66.7% of the total temperature difference (6.0 K). This indicates that the thermal contact resistance arising from lattice mismatch is the dominant thermal resistance of this heterostructure. Figure 4(b) presents the variation in ITC under different structure heights. The ITC is not significantly affected by structural height changes and remains nearly constant at approximately 137.8 $Wm^{-2}K^{-1}$. Therefore, a structure height of 200 nm (i.e., 100 nm for both Si and diamond) is adopted uniformly in subsequent calculations. Compared with GaN/AlN, GaN/SiC, and SiC/diamond interfaces, the ITC of the Si/Damond interface is significantly lower. The reason lies in the substantial mismatch between the phonon spectrum of Si and Diamond. As shown in Figure 4(c), phonons in Si are primarily distributed below 15 THz, whereas in diamond, most phonons are distributed over a broader frequency range of 15–40 THz. Although a limited spectral overlap exists in the 9–15 THz range, this frequency band constitutes only a small portion of the phonons in diamond, resulting in limited elastic phonon transmission across the Si/Diamond interface.

Figure 4(d) illustrates the spectral heat conductance (SHC) distribution of the Si/Damond heterostructure. In bulk Si, the SHC is mainly concentrated in low-frequency acoustic phonons around 5 THz, due to their high group velocities and long mean free paths. A secondary SHC peak appears near 12 THz, corresponding to contributions from optical phonons of Si. In diamond, the SHC spans a broader range of 3–40 THz, with a dominant contribution from acoustic phonons in the 9–15 THz region. A comparison between the SHC near the interface and that in the bulk reveals clear differences. For Si, the contribution of low-frequency acoustic phonons to interfacial heat transport is significantly reduced, while phonons in the 9–15 THz range show enhanced contributions. For diamond, although the dominant heat flux remains in the 9–15 THz range, the contribution from phonons near 15 THz is further enhanced near the interface, while contributions from other frequency ranges are suppressed. These results indicate that the mismatch in phonon distributions and interfacial scattering severely limit elastic phonon transport across the Si/Damond interface, ultimately constraining the interfacial heat transfer performance.

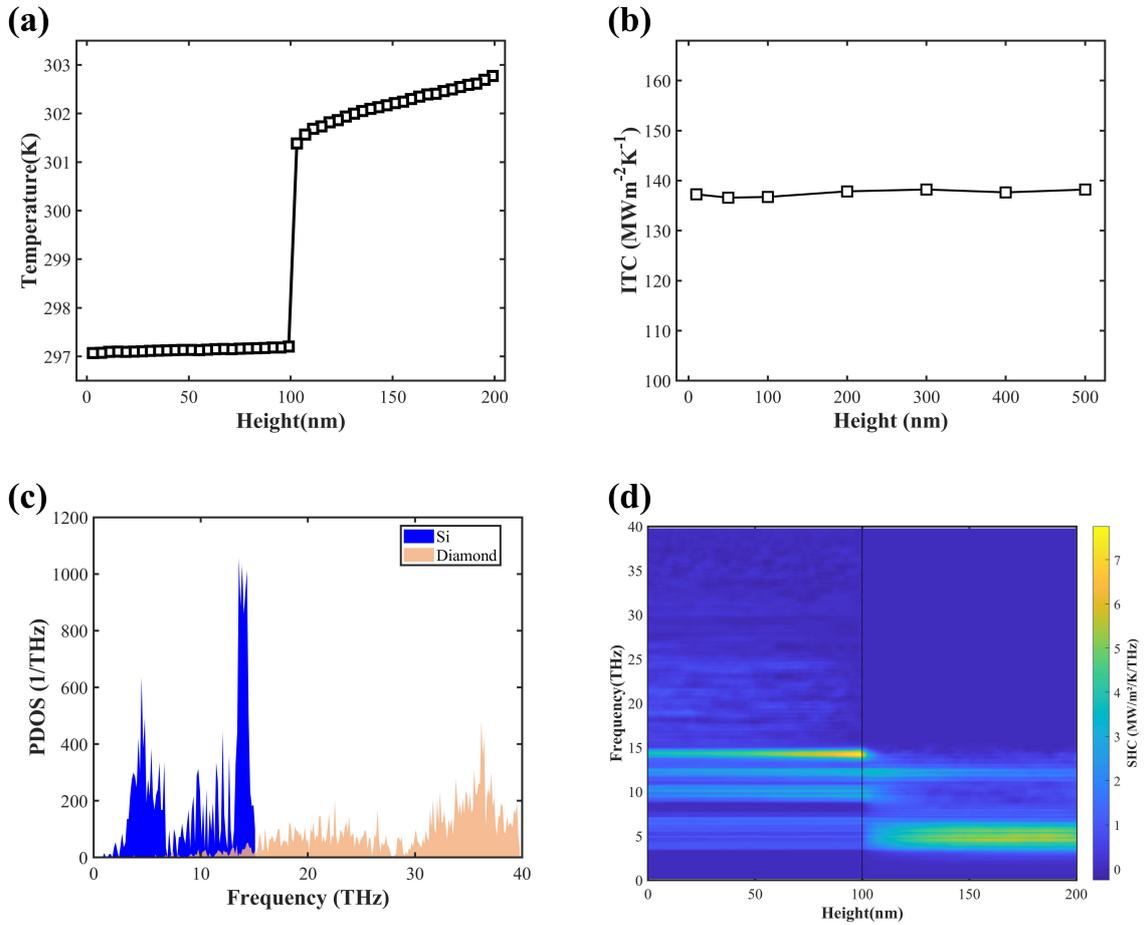

Fig.4. Interfacial thermal transport characteristics of the Si/Damond heterogeneous structure. (a) Steady-state temperature distribution (b) Variation of ITC with different structure heights. (c) Phonon density of states (PDOS) of Si and diamond. (d) Spectral heat conductance distribution.

Figure 5(a) compares the ITC of the Si/Damond interface obtained from different methods. In this work, the full phonon dispersions extracted from first-principles calculations are used to compute the ITC via both the MC method and the DMM method, and the results are compared with available experimental and theoretical data from the literature. The results show that, compared to the simplified DMM approach, which employs Debye densities of states and linear dispersion relations, the ITC calculated using full phonon dispersions is significantly lower. This discrepancy arises because the conventional DMM typically assumes a constant transmission coefficient for all frequencies phonons, simplified as a function of the phonon group velocities of the two materials. As a result, it tends to overestimate the interfacial heat flux. The full-dispersion-based approach allows phonon transmission only when both materials possess corresponding phonon modes at the same frequency, with the transmission coefficient being strongly dependent on the phonon group velocity vector. Figure 5(b) shows the frequency-dependent phonon transmittance from diamond to Si. Due to the substantial mismatch in phonon spectrum, only phonons below 15 THz can be transmitted. Most of the transmittable modes fall within the 9–15 THz range, corresponding to the optical phonons in Si, and the transmittances are relatively low. Phonons above 15 THz cannot contribute to interfacial heat transport due to the lack of available phonon states in Si. This spectral transmittance indicates that interfacial heat is primarily transffferd by phonons in the 9–15 THz range. Consequently, the ITC predicted using full phonon dispersions is lower and more consistent with the best experimental values (approximately 100 $Wm^{-2}K^{-1}$). In contrast, results based on simplified DMM models or empirical-potential molecular dynamics simulations tend to overestimate the ITC.

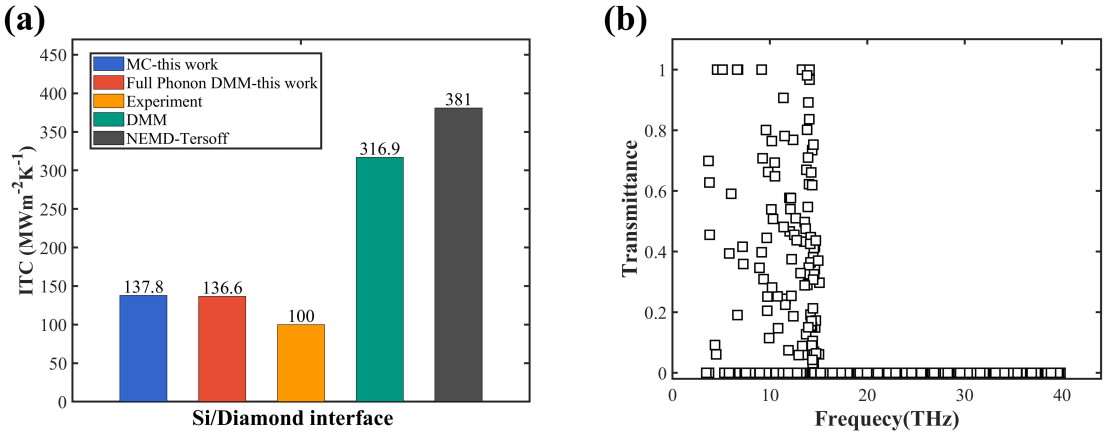

Fig. 5. (a) Comparison of Si/Damond ITC obtained from different methods. Experimental data are taken from Ref. [42], DMM results from Ref. [44], and NEMD simulation results from Ref. [7]. (b) Interface phonon transmittance from diamond to Si.

*3.3 Building the Phonon Bridge for Si/Diamond Interface Using the SiC Interlayer*

Figure 6(a) compares the phonon density of states (PDOS) of Si, diamond, and SiC. Compared to Si, SiC exhibits a broader phonon frequency distribution, with its acoustic phonons spanning 2–18 THz and optical phonons primarily distributed between 21–28 THz. The acoustic phonon spectrum of SiC almost completely overlaps with that of Si and shows substantial overlap with the acoustic phonons of diamond. From the perspective of phonon spectrum matching, SiC is a promising interlayer material capable of forming an effective phonon bridge between Si and diamond to enhance interfacial heat transport. Figures 6(b) and 6(c) present the temperature distributions of the Si/diamond two-layer heterostructure and the Si/SiC/diamond three-layer heterostructure with a 40 nm SiC interlayer. At the Si/diamond interface, the temperature drop is approximately 4.0 K, accounting for 66.7% of the total temperature difference. In contrast, for the Si/40 nm-SiC/diamond heterostructure, the temperature drops at the Si/SiC interface, within the SiC bulk, and at the SiC/diamond interface are 2.57 K, 0.67 K, and 0.31 K, respectively. The total interfacial temperature drop is 3.55 K, 59.2% of the total temperature difference, which indicates that the insertion of the SiC interlayer reduces the overall interfacial thermal resistance.

Figure 6(d) compares the ITC values of the Si/SiC, SiC/diamond, and Si/diamond interfaces, as well as the Si/SiC/diamond structure with a 40 nm SiC interlayer. Benefiting from the better phonon spectral matching, the Si/SiC and SiC/diamond interfaces exhibit significantly higher ITC than the Si/diamond interface. When a 40 nm SiC interlayer is introduced, the Si/SiC/diamond interface achieves an ITC of 202.3 $Wm^{-2}K^{-1}$, representing a 46.6% enhancement compared to the Si/diamond interface. A comparison between the SHC distributions in Figures 5(d) and 6(e) further reveals that the insertion of the SiC interlayer notably increases the SHC contributed by the 3–13 THz range phonons in Si, while a slight reduction is observed in the 13–15 THz. This trend is attributed to larger phonon spectral overlap between Si and SiC in the low-frequency range (3–13 THz), whereas SiC exhibits lower phonon density of states in the 13–15 THz region. On the diamond side of the interface, SHC spans a broad frequency range of 3–28 THz. In particular, the 15–18 THz and 21–28 THz phonons are difficult to access in the Si/diamond interface due to poor spectral overlap. However, after inserting the SiC interlayer, SHC in these bands reaches up to 4 $Wm^{-2}K^{-1}THz^{-1}$. These results demonstrate that the SiC interlayer effectively bridges the phonon spectrum of Si and diamond, forming a phonon bridge that significantly enhances interfacial thermal transport in the heterogeneous structure.

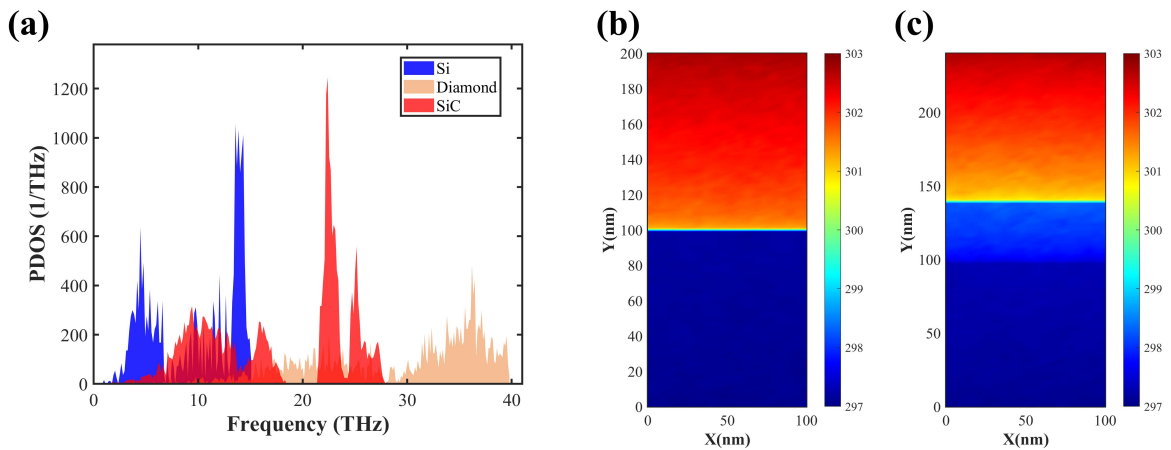

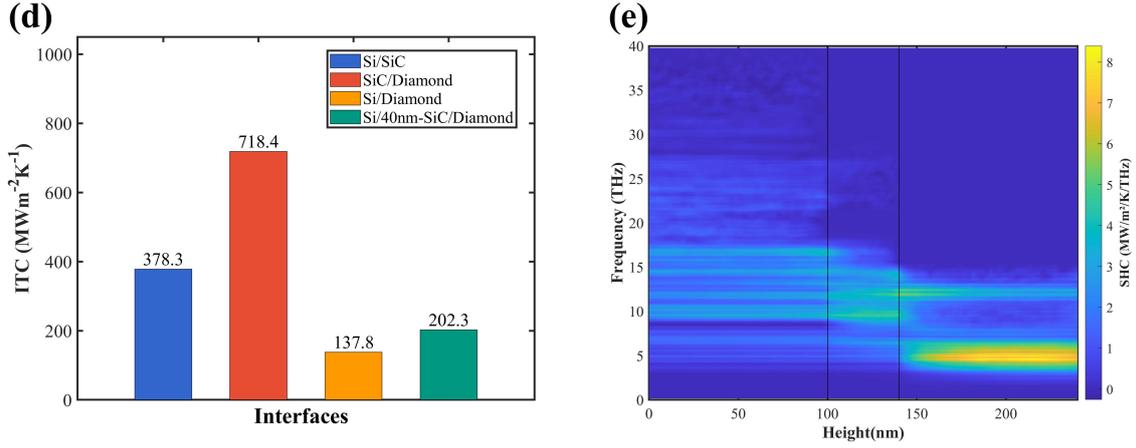

Fig.6. (a) Comparison of PDOS of Si, diamond, and SiC. (b) Temperature distribution of the Si/diamond heterostructure. (c) Temperature distribution of the Si/40 nm-SiC/diamond heterostructure. (d) Comparison of ITC for different interfaces. (e) SHC distribution of the Si/40 nm-SiC/diamond heterostructure.

*3.4 Effect of SiC interlayer thickness*

Figure 7(a) presents the Si/SiC/diamond ITC as a function of SiC interlayer thickness. The results show that ITC increases rapidly with increasing SiC thickness, reaching a maximum value of 202.3 Wm$^{-2}$K$^{-1}$ at the thickness of 40 nm. Beyond this point, further thickening of the interlayer leads to a decline in ITC. When the SiC interlayer is extremely thin, the ITC of the three-layer heterostructure can even fall below that of the Si/diamond heterostructure. For example, at an interlayer thickness of 1 nm, the ITC is only 118.2 Wm$^{-2}$K$^{-1}$, representing a ~14.2% reduction compared to the no-interlayer case. This is primarily attributed to enhanced phonon scattering at the Si/SiC interface. When the interlayer is thin, phonons transmitted from Si into SiC are readily scattered and reflected. Most transmitted phonons are low-frequency acoustic modes with long mean free paths, but the limited SiC thickness provides insufficient space for three-phonon scattering processes that would change the phonon to other modes. As a result, interface thermal transfer remains concentrated in the optical phonons of Si. This explanation can be supported by the SHC distribution shown in Figure 7(b). Under thin interlayer conditions, the SHC is still dominated by Si optical phonons near 12 THz, indicating ineffective coupling to the mid- and high-frequency phonons in diamond. Figure 7(d) shows the thermal resistance contributions of the Si/SiC/diamond heterostructure under different interlayer thicknesses. When the SiC layer is thin, the Si/SiC interface accounts for over 80% of the total interfacial resistance. As the SiC interlayer thickness increases, phonons gain sufficient space within the SiC layer, enabling more frequent three-phonon scattering and frequency conversion, which enhances the bridging effect. When increasing the SiC interlayer thickness, the thermal resistance of the Si/SiC interface decreases rapidly and then stabilizes but raises the bulk thermal resistance of SiC interlayer. As shown in Figures 6(e) and 7(c), when the interlayer thickness is large enough, its bridging function saturates, and the thermal resistance at the Si/SiC and SiC/diamond interfaces no longer changes significantly. Therefore, excessive interlayer thickness leads to a net reduction in overall ITC due to increased interlayer bulk resistance. In summary, the optimal interlayer thickness results from a trade-off between the phonon-bridging effect and the added bulk thermal resistance of the interlayer. This balance determines the maximum achievable ITC for the three-layer heterostructure.

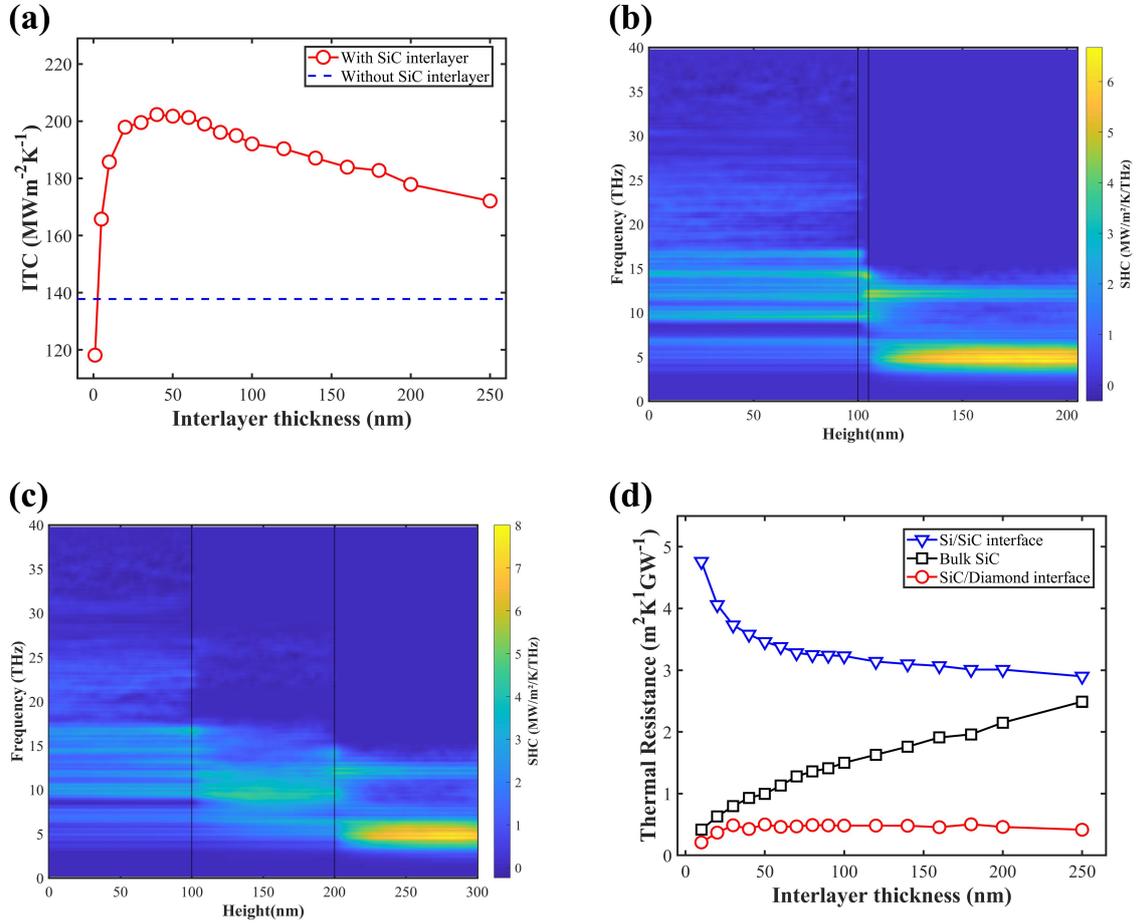

Fig.7. (a) Variation of ITC with different SiC interlayer thicknesses. (b) SHC distribution for the Si/5 nm-SiC/diamond heterostructure. (c) SHC distribution for the Si/100 nm-SiC/diamond heterostructure. (d) Variation of three thermal resistance components in the Si/SiC/diamond structure as a function of interlayer thickness.

*3.5 Comparison of interlayers with different thicknesses and materials*

The ITC of Si/interlayer/Diamond heterostructures incorporating thirteen different interlayers with varying thicknesses, including SiC, AlN, α-$Si_3N_4$, β-$Si_3N_4$, and $Al_xGa_{1-x}N$ (x ranges from 0.1 to 0.9), are compared, and the results are presented in Figure 8. The results show that for SiC, AlN, α-$Si_3N_4$, β-$Si_3N_4$, and $Al_{0.1}Ga_{0.9}N$, ITC initially increases with interlayer thickness and then decreases, indicating the existence of an optimal thickness that maximizes ITC. In contrast, for $Al_xGa_{1-x}N$ with *x* ranging from 0.2 to 0.9, ITC decreases with increasing interlayer thickness. The reason is that materials like SiC and AlN have high thermal conductivities and long phonon mean free paths. When the interlayer thicknesse is small, transmitted phonons can more easily reach and scatter back from the second interface, causing high interfacial resistance at the Si/interlayer boundary. As the interlayer thickness increases, the Si/interlayer interfacial resistance decreases rapidly due to enhanced phonon redistribution, but the bulk thermal resistance of the interlayer increases, resulting in a non-monotonic dependence and an optimal thickness. For $Al_xGa_{1-x}N$ (x rangs from 0.2 to 0.9), the low thermal conductivity and short phonon mean free paths make the bulk thermal resistance more sensitive to thickness than the interfacial resistance. As a result, thinner interlayers minimize bulk resistance and yield higher ITC. The optimal thicknesses for SiC, AlN, α-$Si_3N_4$, β-$Si_3N_4$, and $Al_{0.1}Ga_{0.9}N$ are 40 nm, 40 nm, 30 nm, 10 nm, and 5 nm, respectively. For $Al_xGa_{1-x}N$ (x rangs from 0.2 to 0.9), the optimal thickness is 1 nm. As shown in Figure 8(c), the ITC values at the respective optimal interlayer thicknesses indicate that all interlayers except α-$Si_3N_4$ can enhance interfacial heat transport across the Si/diamond interface. Among them, SiC yields the highest improvement, with a 46.6% increase in ITC, followed by AlN with a 21.9% improvement. $Al_{0.1}Ga_{0.9}N$ also exhibits a substantial enhancement (21.3%) due to its phonon spectrum being like that of AlN.

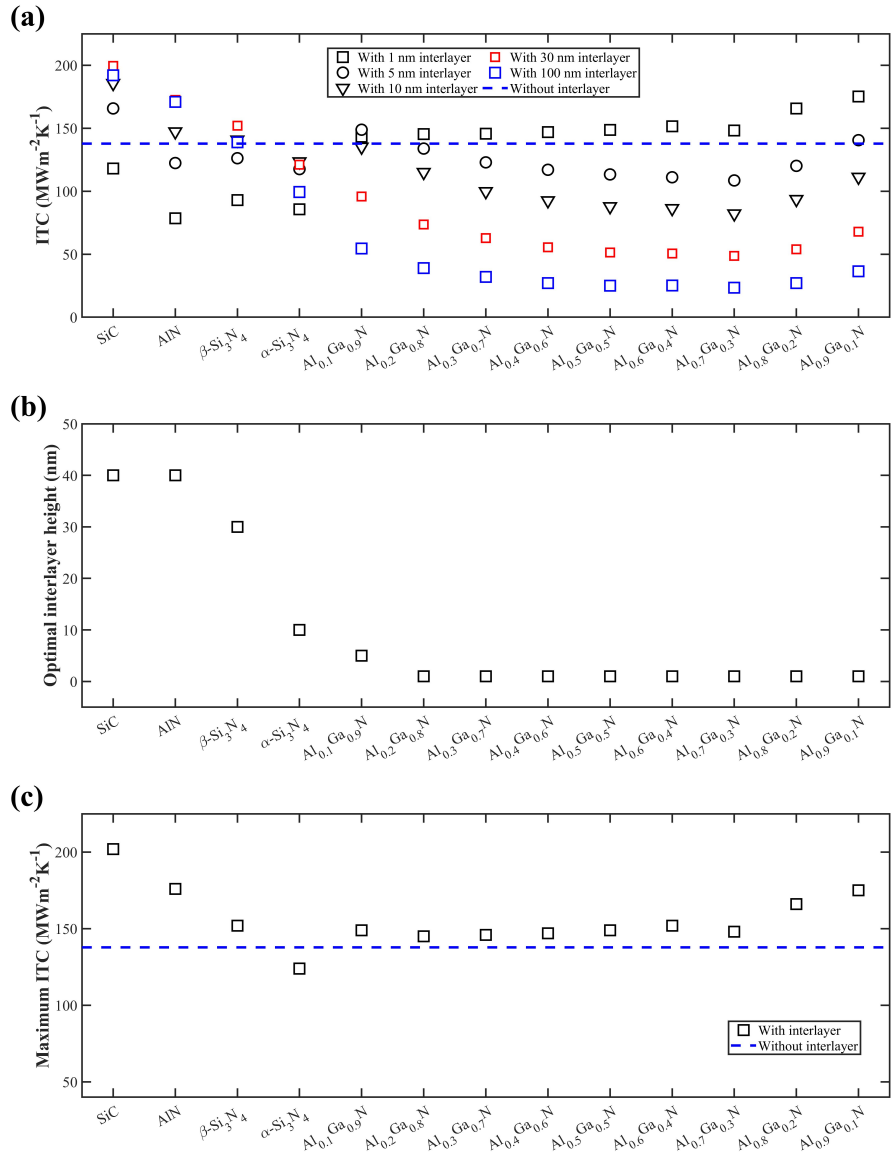

Fig.8. (a) Effect of interlayer material and thickness on the ITC of the Si/interlayer/diamond heterostructure. (b)

Optimal interlayer thickness for different interlayer materials. (c) Maximum ITC for different interlayer materials.

**4 Conclusions**

A heat transfer model for three-layer heterostructures is developed by combining First-principles calculations with the Monte Carlo method. The mechanism of enhancing interfacial thermal transport performance in Si/Diamond heterostructures using phonon bridge is revealed. The main conclusions are as follows:

(1) The significant phonon spectrum mismatch between Si and diamond limits heat transfer at the interface, resulting in an interface thermal conductance of only 137.8 $Wm^{-2}K^{-1}$.

(2) The SiC interlayer effectively bridges the phonon spectrum of Si and diamond, forming a phonon bridge that significantly enhances interfacial thermal transport. When a 40 nm SiC interlayer is introduced, the three-layer heterostructure achieves an ITC of 202.3 $Wm^{-2}K^{-1}$, representing a 46.6% enhancement compared to the Si/diamond two-layer heterostructure.

(3) When the interlayer is thin, the phonon bridge enhancement effect is suppressed, while a thick interlayer will increase the interlayer bulk thermal resistance. For the SiC interlayer, the optimal thickness is 40 nm.

(4) Among the thirteen interlayers studied, including SiC, AlN, α-$Si_3N_4$, β-$Si_3N_4$, and $Al_xGa_{1-x}N$ (x ranges from 0.1 to 0.9), SiC showed the best enhanced effect, followed by AlN.

These findings provide essential insights into the phonon bridge mechanism at heterogeneous interface thermal transport and offer valuable theoretical guidance for designing heterostructures with enhanced thermal transport performance.

**Acknowledgments**

This work was supported by the National Natural Science Foundation of China (Grant NO. 52006102), the Fundamental Research Funds for the Central Universities (No.30923010917).